# Measurement of electron temperature in a non-equilibrium discharge of atmospheric pressure supported by focused microwave radiation from a 24 GHz gyrotron

Sergey Sintsov, Alexander Vodopyanov, Dmitry Mansfeld

___________

S.V. Sintsov *, A.V. Vodopyanov, D.A. Mansfeld

Institute of Applied Physics RAS, 46 Ul'yanov Street, 603950, Nizhny Novgorod, Russia

sins@ipfran.ru *

___________

Abstract

A microwave discharge of atmospheric pressure, maintained by focused CW radiation of a 24-GHz gyrotron in an argon flow in an external air atmosphere, was investigated. The electron temperature was determined by the current-voltage curve of a dual Langmuir probe placed in a plasma torch. The electron temperature was also estimated from plasma emission spectra within the framework of a coronal plasma model. The obtained values of the electron temperature coincide within the measurement accuracy. Also, the electron temperature is many times higher than the gas temperature. This fact allows us to stand of a significantly non-equilibrium atmospheric pressure plasma.

## 1  Introduction

Currently, there is a growing interest in non-equilibrium atmospheric plasma sources for a wide range of plasma-chemical applications.[1] This is due to the requirements of modern industry for the technologies with highly stable molecules decomposition and processes with a high activation threshold. Also, the implementation of such a non-equilibrium discharge at atmospheric pressure is highly required for high volume production in industry.[2] A high specific energy input is required to ensure conditions for effective heating of the electron component and to maintain significantly non-equilibrium atmospheric pressure plasma. This can be realized using gyrotrons - the powerful sources of microwave radiation.[3–6] The discharge maintained in quasi-optical wave beams is ideal for obtaining and practical use of reactive plasma with a high purity due to the absence of electrodes and the localization of the discharge far from the chamber walls.[7]–[9]

There are many papers devoted to the study of various types of microwave discharges, sustained by electromagnetic waves in a wide range of wavelengths and pressures. The heating radiation wavelength is from the centimeter to millimeter range, and the pressure range is from $10^{-6}$ to 760 Torr. Discharges of this type is practically used. For plasma technologies of etching and deposition of thin-films and functional coatings,[10]–[14] effective sources of vacuum ultraviolet light and soft X-rays.[15] However, these applications are often implemented in pulsed low-pressure microwave discharges (less than $10^{-2}$ Torr), including conditions of electron cyclotron resonance heating. Low pressure operating is often a limitation for volume production plasma chemistry.

The studies of the plasma at sub-atmospheric and atmospheric pressure supported by the continuous microwave radiation with frequency of 2.45 GHz are widely known.[16], [17] Research interest is concerned with the wide availability of such microwave sources and the relatively simple electrodeless design of the plasma torch. Plasma supported by magnetron radiation is widely used for carrying out chemical reactions in the pressure range from $10^{-3}$ to 760 Torr.[18] However, the power density currently available in these sources is insufficient to maintain a non-equilibrium discharge at atmospheric pressure due to the relatively long wavelength of radiation and consequently small power load to the plasma. So, the E / N parameter is quite low which results in thermal equilibrium plasma.

In this paper, we have studied the discharge at atmospheric pressure, sustained by focused radiation of a gyrotron with the frequency of 24 GHz and power of up to 5 kW. The use of such an expensive source due to its unique properties. The radiation wavelength of the used gyrotron is 10 times lower than the radiation wavelength of a commercial magnetron. This makes it possible to achieve large values of the radiation power density when it is focused in the quasi-optical path. Also for industrial applications, there are a number of gyrotron complexes that have an output power in a continuous mode, unattainable for modern magnetrons. In this work, the electron temperature was measured by two independent methods: by the emission spectra of the plasma and by the voltage-current characteristic of the double probe. The study of such discharge is not only interest as a new object in plasma physics. Also it may find a wide range of applications in chemistry due to the non-equilibrium plasma.

## 2     Experimental Section

### 2.1     Experimental Techniques

In this work, the technological gyrotron with the radiation frequency of 24 GHz and output power of up to 5 kW in the continuous mode was used as a source of microwave radiation.[4], [19]     The experimental setup is shown in Figure 1. The generated radiation along a circular waveguide (1 in Figure 1) with a diameter of 32.6 mm was converted to a Gaussian beam with linear polarization. The radiation is introduced into the gas-discharge chamber through a cooled window made of boron nitride (2 in Figure 1). The parabolic mirror (3 in Figure 1) was used to increase the energy flux density. The microwave beam minimal diameter was 1.5 cm at a level of 0.5 $E^2_{max}$. The radiation intensity in this area reached 5 kW/cm$^2$ at a maximum output power of gyrotron. This corresponds to intensity of the RMS electric field of 1.9 kV/cm. The end of the metal tube (4 in Figure 1) was led to the beam waist region. This tube was used to supply the plasma-forming gas. The inner diameter of the tube was 4 mm. At the end of the gas tube, discharge (5 in Figure 1) was initiated by the spark discharge and continuously supported by focused microwave radiation. The discharge is an elongated torch, with the length of 2 to 4 cm and a diameter equal to the diameter of the gas tube (Figure 2). Argon was used as a plasma forming gas, and the discharge was ignited in surrounding air at atmospheric pressure. The flow rate of argon could be varied from 5 to 30 l/min. The torch length changed proportion only to the microwave power and did not depend on the gas flow rate (Figure 3).

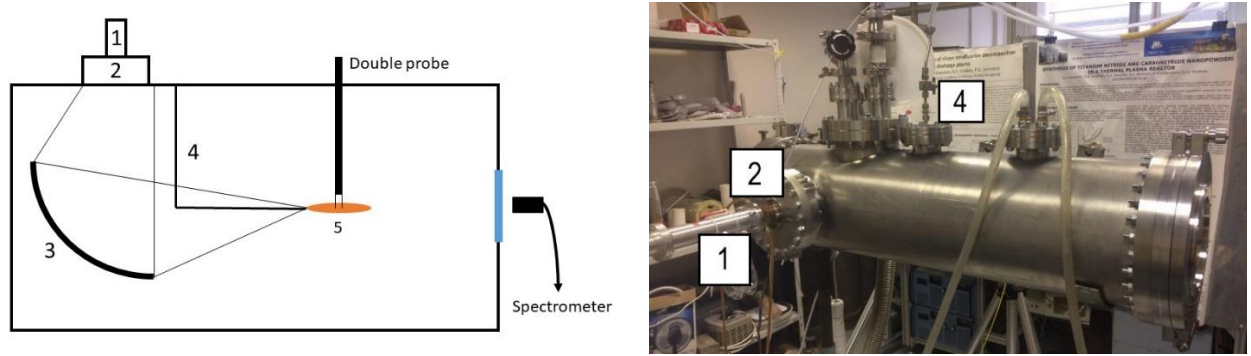

*Figure 1. Scheme and photo of the installation. 1. Circular waveguide; 2. Microwave window; 3. Parabolic mirror; 4. Tube suppling plasma forming gas; 5. Plasma torch.*

A water circuit was placed as a passive load for the supplied microwave power inside the gas-discharge chamber in the area behind the focal waist. The power deposited in the torch was calculated by measuring the input power to the chamber and the power absorbed in this water circuit.

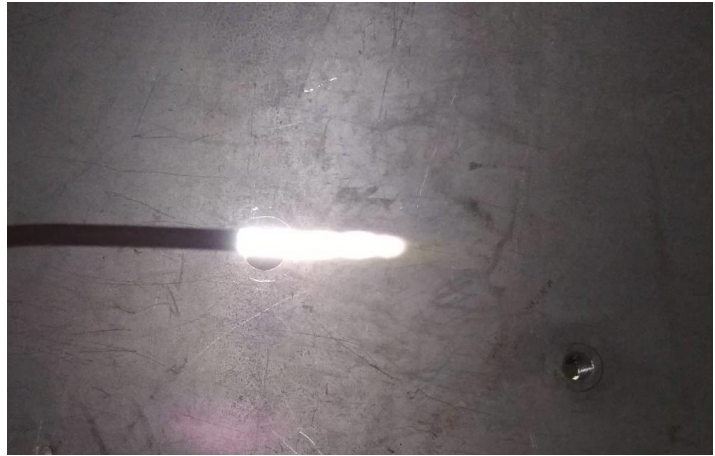

*Figure 2. Photo of a torch supported by focused microwave radiation from. Electromagnetic wave propagates from left to right. Exposure time 0.1 s.*

It varies in the range from 100 to 500 W, which corresponds to the 10% of the power introduced into the chamber. Such a small absorption coefficient is due to the fact that the area of the microwave beam in the waist is about 10 times larger than the cross-sectional area of the torch. Thus, most of the microwave power passes the discharge without interaction. The absorption coefficient can be increased either by increasing the diameter of the gas stream, or by reducing the wavelength of the heating field. Taking into account the fraction of the power absorbed by the torch, the gas temperature can be estimated from above by the specific enthalpy of heating of argon. Assuming that all the absorbed power eventually goes into thermal heating of the plasma, the equilibrium temperature of the torch does not exceed 500 K at the maximum value of the input power and the minimum argon flow rate. It also can be concluded that the gas temperature is low by the absence of traces of melting of the copper gas tube and the double probe placed in the plasma.

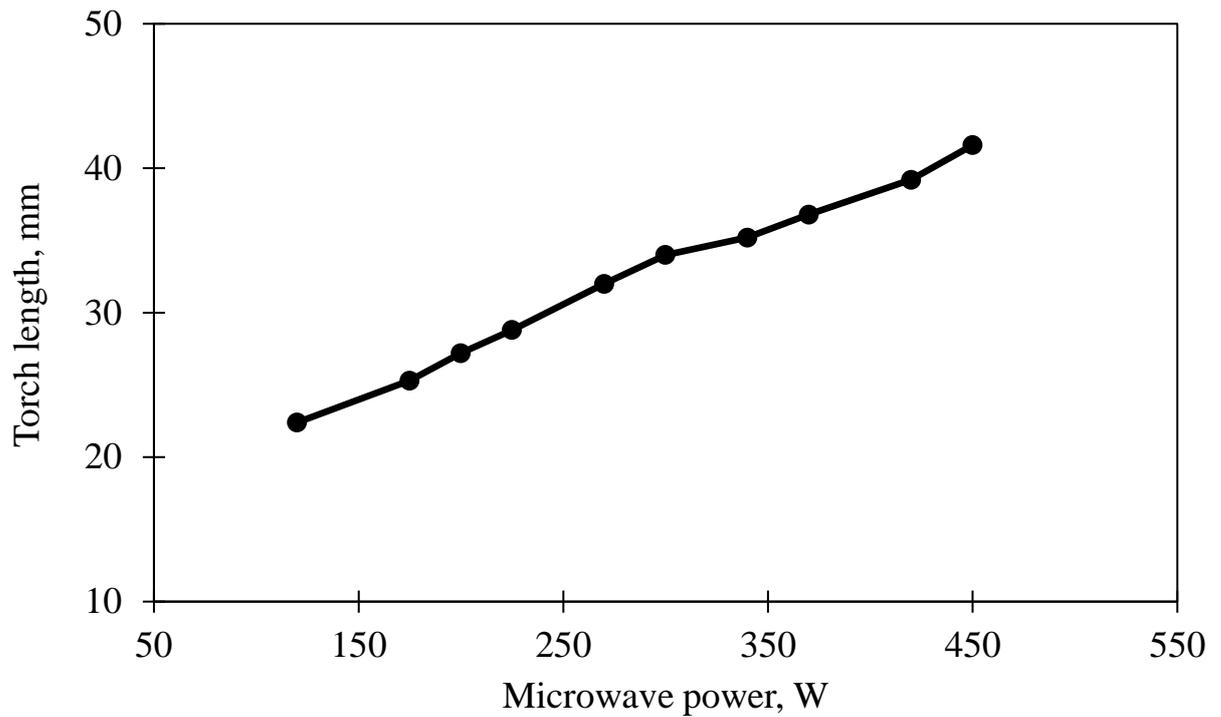

*Figure 3. Torch length dependence on absorbed power. Flow rate of Ar is 15 l/min.*

The results of the electron temperature measurements by two independent methods are below. The electron temperature was measured by the voltage-current characteristic of a double probe placed in a plasma and by emission lines of argon of the visible wavelength range.

## 2.2 Probe Measurements

Significant limitations in the use of probe methods for studying atmospheric pressure plasma are concerned with the energy flow that the probe can withstand. The plasma investigated in this experiment is non-thermal, which made it possible to successfully perform probe measurements.

However, high gas pressure can affect the accuracy of measuring the electron temperature in the case the electrons are cooled strongly when they hit the probe. In this case, the measured electron temperature may differ from the temperature in the plasma volume.[20] This cooling should be taken into account if Equation 1 is fulfilled:

$$\frac{r_{probe}}{\lambda_{en}} >> \sqrt{\frac{M_i}{m_e}} \tag{1}$$

where $r_{probe}$ is the probe radius, $\lambda_{en}$ is the electron mean free path between electron-neutral collisions, $m_e$ is the electron mass, $M_i$ is the mass of argon ion.

Then, for a symmetric dual probe,[20] the electron temperature can be obtained by Equation 2. Temperature is expressed in eV.

$$T_e = \frac{I_{sat}}{6.16}\left(\frac{dV_p}{dI_p}\right)_{I_p=0} \tag{2}$$

where $I_{sat}$ is the ionic saturation current in amperes, $dV_p/dI_p$ is the slope of the current – voltage characteristic at zero current. For the plasma under this study, Equation 1 is fulfilled, since the electron mean free path is much less than the probe size. Therefore, when processing the current – voltage characteristics, the electron temperature was estimated using Equation 2. Also, it is possible to estimate the value of the electron density on the current-voltage characteristic. However, the obtained values can be strongly distorted due to the large number of collisions in the double layer near the probe surface for atmospheric pressure discharges.[21] Therefore, the measurement of electron density was not carried out in this work.

Figure 4 shows the measurement scheme of the current – voltage characteristics and a photo of the used double probe. Two tungsten wires with a diameter of 100 μm were enclosed in quartz insulation without mechanical contact with each other. The length of the wires in contact with the plasma was 1 mm.

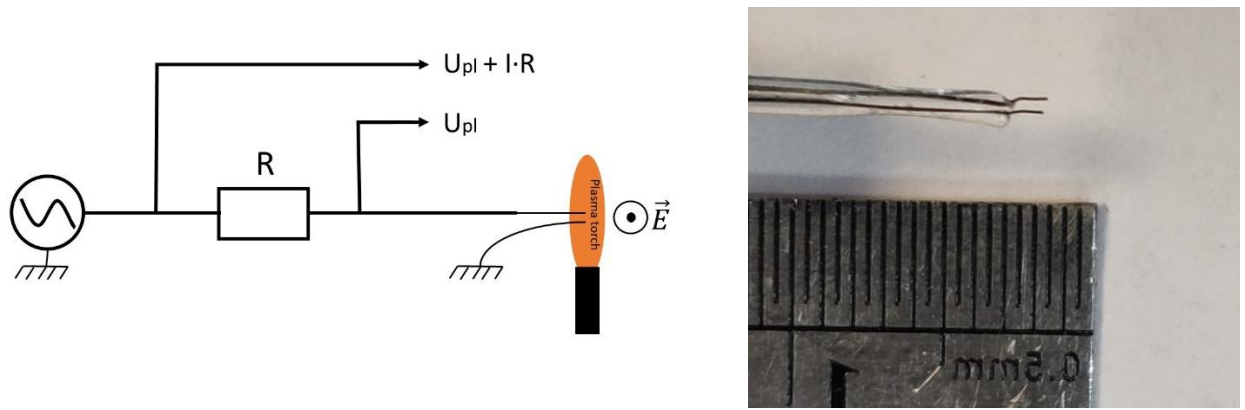

*Figure 4. The scheme of probe measurements and photo of a used Langmuir double probe.*

The probe could be heated by microwave radiation, that is why it was made of tungsten. Quartz was chosen as a dielectric because it is a refractory insulator with a small tangent of angular loss in the microwave wavelength range. The probe was introduced into the chamber from the side, perpendicular to the torch and perpendicular to the external electric field to minimize interference. All measurements were taken in the center of the body of the torch. A sinusoidal voltage generator was used as the source for the probe circuit. For a single series of measurements, several tens of voltage periods were taken, which were subsequently averaged. The exposure time of the probe in the plasma did not exceed 10 s. A typical current-voltage characteristic of the probe is shown in Figure 5. The electron temperature was derived from the characteristics using Equation 2. The rotation of the probe placed in the plasma around its axis did not affect the parameters of the current – voltage characteristics. This indicates that the plasma is isotropic in the region under study.

The measurements were carried out various microwave power and argon flow rate. The measurement error was determined as the statistical variation of the obtained temperature value in each operation mode of the plasma torch, and have not exceed 0.3 eV. Figure 6 shows the dependence of the electron temperature on the power absorbed by the torch at different argon flow rates. The electron temperature was about 1 eV.

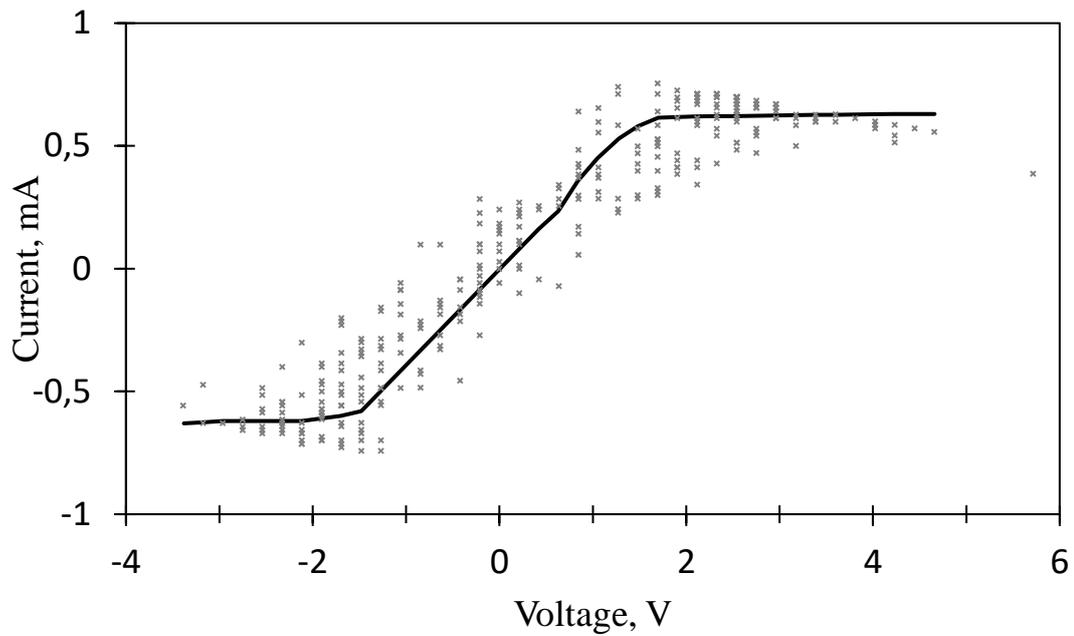

*Figure 5. Averaged volt-ampere characteristic of the probe. Heating power 1.5 kW, argon flow rate 10 l/min.*

The electron temperature value does not depend on the heating power and argon flow rate within the measurement accuracy. An increase in the power of microwaves results in increase of the torch length. It should be noted that the obtained value of the electron temperature is many times higher than the estimated gas temperature. This indicates that the non-equilibrium discharge was realized.

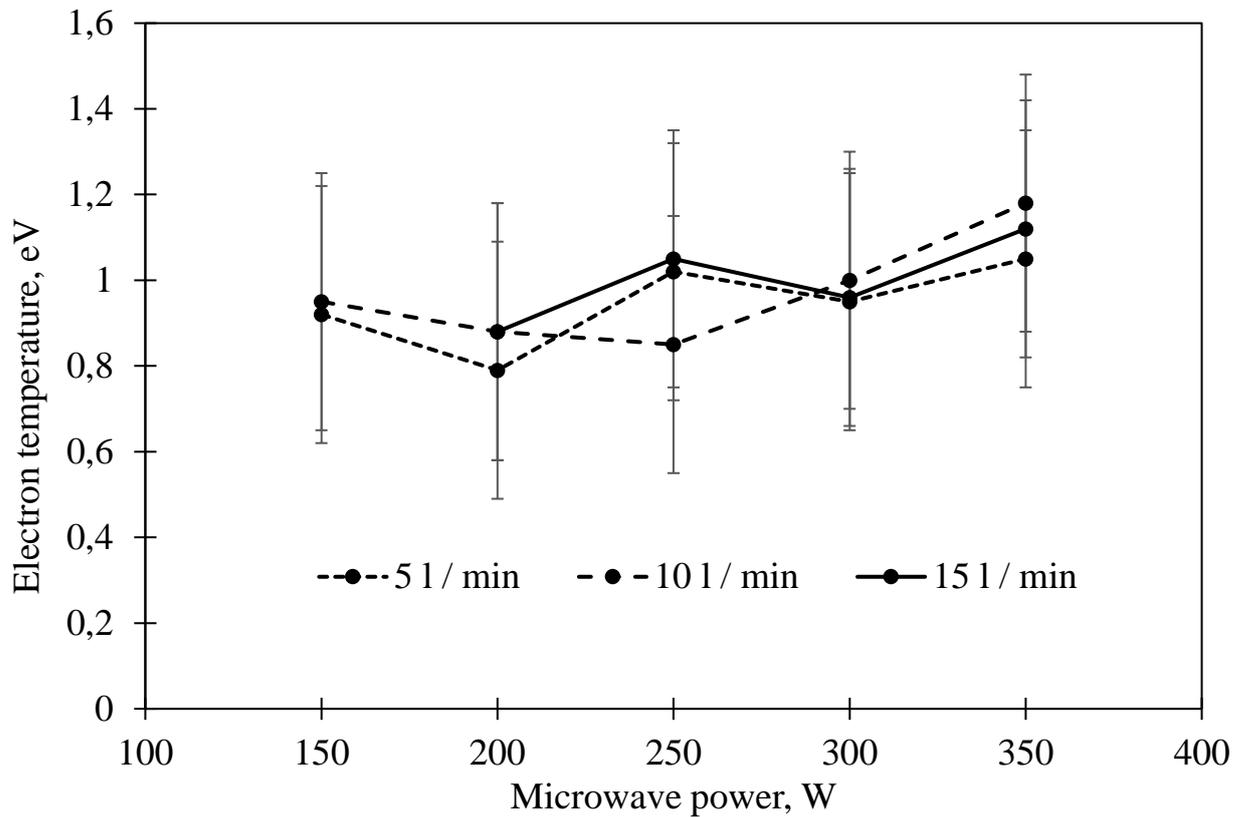

*Figure 6. The dependence of the electron temperature, measured by double probe, on the absorbed power at different argon flow rates.*

## 2.3 Measurement of Emission Spectra

The emission spectra of the torch were measured to verify the measured electron temperature using double probe. The spectrograph MS5204i (SOL instruments) used to record emission spectra in the range of 300–1000 nm with a resolution of 0.03 nm. The radiation was collected from the spot of 2 mm in diameter from the central part of the torch. This was ensured by focusing the radiation from the desired region by a system of optical lenses. Figure 7 shows the typical emission spectrum of the plasma under study. It can be seen that emission lines dominate continuum emission, this partly confirms the assumption of the non-equilibrium of this type of discharge. The emission spectra of the torch were obtained at different values of heating power and argon flow rates. The electron temperature was evaluated for each mode.

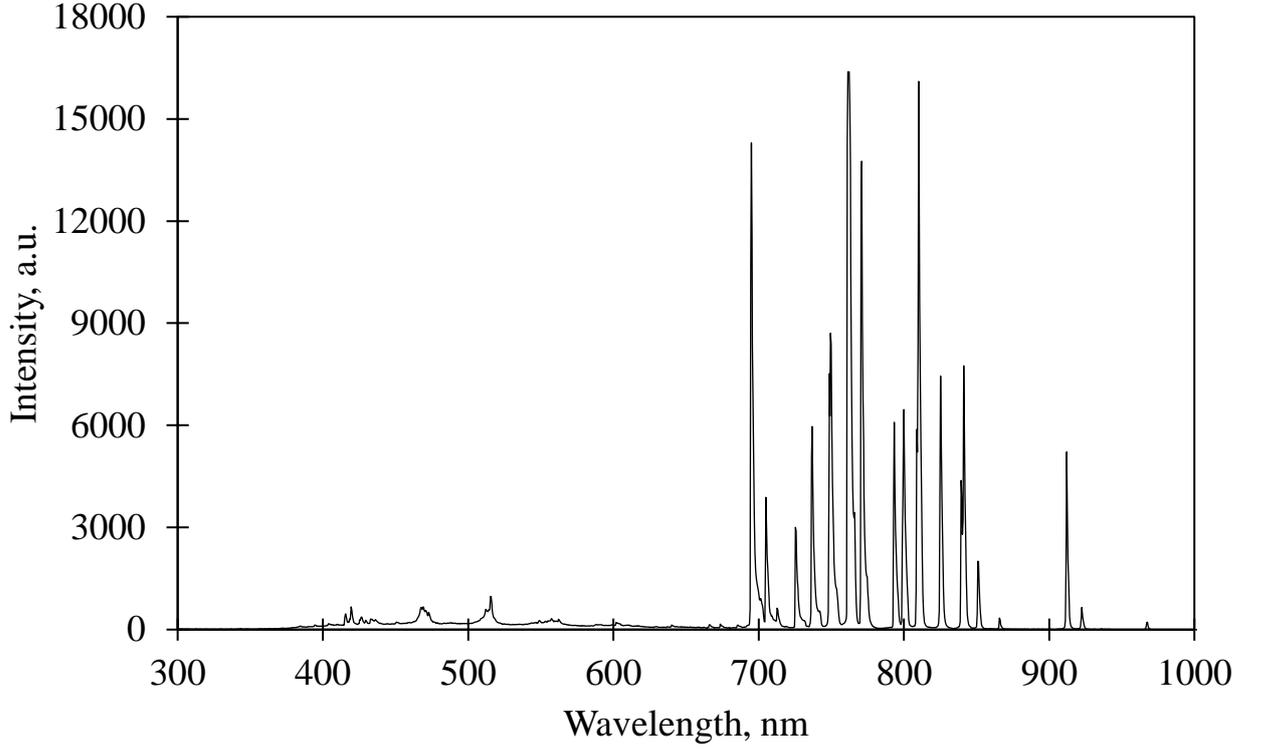

*Figure 7. The characteristic emission spectrum of the argon torch of atmospheric pressure. Heating power 1.5 kW, argon flow rate 10 l/min.*

The electron temperature was estimated using the coronal plasma model.[22] The emission transition lines of the argon atom from the 2p level to the 1s level with the highest Einstein coefficients were selected for this. It is known that the population distribution of these levels is well described by the coronal model even in an atmospheric-pressure collisional plasma, where the frequency of collisions of electrons with neutrals is much higher than the frequency of the external field.[23] The argon lines used for the electron temperature estimations are presented in Table 1.

One can determine the temperature of the electrons using Equation 3 using measured relative intensities of the two emission lines and taking into account the atomic transition constants. [22,23]

$$\frac{I_{ki}}{I_{mn}} = \frac{A_{ki} g_k \lambda_{mn}}{A_{mn} g_m \lambda_{ki}} \exp\left(-\frac{E_k - E_m}{kT_e}\right) \qquad (3)$$

where I is the relative intensity; A is the Einstein coefficient; λ is the wavelength for optical transitions k → i and m → n; g and E are respectively the statistical weight and binding energies of the upper levels of k and m; $T_e$ is the electron temperature.

*Table 1. Optical transitions between the 1s and 2p levels of the argon atom*

| Transition | Einstein Coefficient [s$^{-1}$] | Wavelength [nm] |
|---|---|---|
| Ar(2p1) → Ar(1s2) + hν | $4.5 \times 10^7$ | 750.4 |
| Ar(2p2) → Ar(1s2) + hν | $1.5 \times 10^7$ | 826.5 |
| Ar(2p3) → Ar(1s2) + hν | $2.2 \times 10^7$ | 840.8 |
| Ar(2p4) → Ar(1s2) + hν | $1.4 \times 10^7$ | 852.1 |
| Ar(2p5) → Ar(1s4) + hν | $4.0 \times 10^7$ | 751.5 |
| Ar(2p6) → Ar(1s5) + hν | $2.5 \times 10^7$ | 763.5 |

The electron temperature was determined by the method described above for each operation mode of the plasma torch. $T_e$ was determined for each pair of lines from Table 1. Then, the obtained values are averaged, and the standard deviation determines the measurement error. Figure 8 shows the dependences of the electron temperature on the absorbed power at different argon flow rates. It can be seen that, the electron temperature is about 1.2 ± 0.4 eV and does not change with the power within the measurement accuracy.

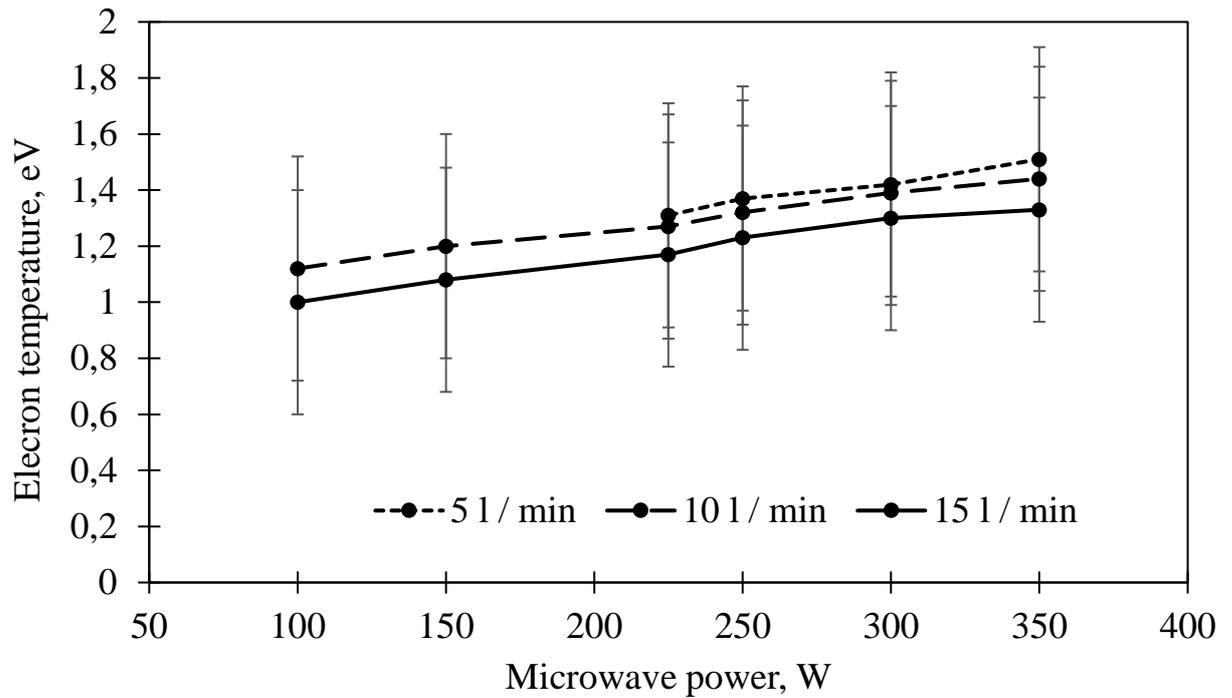

*Figure 8. The dependence of the electron temperature measured by argon emission transitions on the absorbed power at different argon flow rates.*

## 3    Results and Discussion

The electron temperature measured by probe and optical emission spectroscopy methods coincide within the limits of error. This confirms the reliability of both used diagnostic methods. In order of magnitude, the electron temperature was 1 eV, which significantly exceeds the estimated gas temperature, which is at a level of 0.04 eV. This indicates the non-equilibrium of the plasma, supported by focused microwave radiation at atmospheric pressure. Within the limits of measurement accuracy, the value of the electron temperature did not change with an increase in the power absorbed by the discharge and an increase in the argon flow rate. An increase in the power supplied to the discharge only led to a proportional increase in the length of the torch (Figure 3). The open end of the gas feeding tube was located in the plane of the microwave beam waist. So, the microwave field strength decreases with the distance from the tube end. The microwave field reaches the value sufficient to maintain the discharge farther and farther away from the waist area with an increase in the incident microwave power, which result in the increasing torch length.

The fact that electron temperature does not depend on microwave power may indicate that the electron density in the discharge has reached a cut-off density ($7.1 \cdot 10^{12}$ cm$^{-3}$) for the frequency of the heating radiation. An increase in the field strength at the point where the breakdown conditions had already been fulfilled and the plasma existed did not lead to an increase in the electron temperature. This may be due to the redistribution of the field due to the refraction of the heating electromagnetic wave on the plasma object.

Acknowledgments: The authors are grateful to prof. Stepanov A.N. for the discussion of the work and valuable comments. This work was supported by the Russian Foundation for Basic Research, projects No. 17-02-00785 and No. 18-29-21014.

Keywords: microwave discharge, non-equilibrium plasma, gyrotron, plasma chemistry.